# A Boundary Thickening-based Direct Forcing Immersed Boundary Method for Fully Resolved Simulation of Particle-laden Flows

Maoqiang Jiang, Zhaohui Liu†

*State Key Laboratory of Coal Combustion, School of Energy and Power Engineering,*

*Huazhong University of Science and Technology, Wuhan 430074, PR China*

† Corresponding author. Tel.: +86-027-8754-2417-8308. E-mail address: zliu@hust.edu.cn (Z. Liu).

**Abstract:**

A boundary thickening-based direct forcing (BTDF) immersed boundary (IB) method is proposed for fully resolved simulation of incompressible viscous flows laden with finite size particles. By slightly thickening the boundary thickness, the local communication between the Lagrangian points on the solid boundary and their neighboring fluid Eulerian grids is improved, based on an implicit direct forcing (IDF) approach and a partition-of-unity condition for the regularized delta function. This strategy yields a simple, yet much better imposition of the no-slip and no-penetration boundary conditions than the conventional direct forcing (DF) technique. In particular, the present BTDF method can achieve a numerical accuracy comparable with other representative improved methods, such as multi-direct forcing (MDF), implicit velocity correction (IVC) and the reproducing kernel particle method (RKPM), while its computation cost remains much lower and nearly equivalent to the conventional DF scheme. The dependence of the optimum thickness value of boundary thickening on the form of the regularized delta functions is also revealed. By coupling the lattice Boltzmann method (LBM) with BTDF-IB, the robustness of the present BTDF IB method is demonstrated using numerical simulations of creeping flow (Re=0.1), steady vortex separating flow (Re=40) and unsteady vortex shedding flow (Re=200) around a circular cylinder.

## 1. Introduction

Particle-laden flows are involved in a vast number of process systems ranging from chemical and petrochemical processes to energy conversion, environmental flows and biomedical flows, etc. Hence, accurate and efficient numerical simulation of these flows is of substantial importance for fundamental research, which includes understanding and optimizing these complex flow problems. Fully (or particle) resolved direct numerical simulation (F/PR-DNS), as a first-principles approach to developing accurate models for particle-laden flows at all levels of statistical closure by solving the governing fluid equations with exact boundary conditions imposed at each particle surface[1], has rapidly become a focus in fluids research in recent years [1-3]. F/PR-DNS is the most accurate technique for revealing the dynamic fundamentals of interactions between the continuum fluid phase and the suspended particle phase, which cannot be implemented in experiment or other numerical techniques such as the two-fluid model or point particle model.

Over the last few years, the immersed boundary method (IBM) [4, 5] has become a popular method for FR-DNS of particle-laden flows [3] with a huge number of finite size particles. It was first presented by Peskin [6] in the 1970s for simulation of blood flow with flexible valves inside the heart. The key feature for this method is that it decouples the solution of the flow field from the implementation of the boundary condition and is able to apply a simple Cartesian mesh for flows with moving complex boundaries. In addition, the tedious re-meshing process for moving geometries at each time step, which is commonly required in the arbitrary Lagrangian–Eulerian (ALE) approach, is also removed. On the other hand, the fluid is solved by the Navier–Stokes equation (NSE) or lattice Boltzmann equation (LBE), both outside and inside the particles, while enforcing a fictitious force density on the local nearby fluid to achieve no-slip conditions on each particle's boundary.

While the general idea is simple, the challenge and a topic in current research is to develop an IBM that is both sufficiently accurate and computationally efficient [3]. The original IBM pioneered by Peskin [4, 6] is a continuous forcing or feedback forcing approach. The boundary force density defined as a singular function along the boundary was computed by Hooke's law with an appropriate stiffness parameter, and then included feedback to the neighboring Eulerian grids with a regularized delta function. The feedback forcing approach is simple and straightforward, but it suffers from a severe time step restriction in numerical stability due to the preconditioned one or two stiffness parameters that need to be tuned according to the flow conditions and particle properties. This significantly reduces the efficiency, especially for an unsteady flow and/or rigid particles.

The major breakthrough was initiated by Uhlmann [7] in 2005 who combined the regularized delta function approach with the direct-forcing (DF) approach developed by Mohd-Yusof [8] and Fadlun et al.[9]. In this way, the numerical efficiency was remarkably enhanced, while the numerical non-smoothing problem (as the grid-locking phenomenon) was remarkably suppressed. The immersed boundary (IB) force density is directly computed by the difference between the unforced fluid velocity and the particles' boundary velocity on the Lagrangian points. The hydrodynamic forces and torques acting on particles can be obtained by simply integrating the IB force density distributed on the Lagrangian points.

It should be pointed out that the method of direct forcing immersed boundary (DF-IB) coupled with NSE or with LBM were two approaches that developed alongside each other, as shown in Table 1. The main difference between NSE and LBM is the pressure term solver. The former uses a complicated projection approach [10, 11] while the latter uses a simple explicit method [12]. Accordingly, the computational step framework follows a prediction, forcing and projection model in NSNSE while the framework is a prediction, forcing, collision and stream model in IB-LBM. Feng and Michaelides [13] first combined the DF-IB method with LBM to solve 3D particulate flow problems. However, the IB force density was evaluated by NSE in their study. Dupuis et al. [14] proposed a pure and the simplest DF IB-LBM where both the IB force density and fluid flow were solved by the LBE.

Because the Eulerian grids inside the fluid and the Lagrangian points on the particle's surface do not overlap, the smooth regularized delta function is used in velocity interpolation from the former to the latter, and the force density spreads from the latter back to the former, either in the DF IB-NSE [7] or in the DF

IB-LBM [14]. This interface smoothing technique presents an important advantage in that it suppresses undesired high-frequency oscillations in the force and torque acting on particles when the particles move over the Eulerian grids. However, it still possesses a major disadvantage. That is, the no-slip and no-penetration boundary conditions cannot be strictly fulfilled in the converged state, because the force density cannot be fully reconstructed after interpolation and spreading operations. One of the reasons for this is that the Lagrangian points may "share" the Eulerian points during the Eulerian–Lagrangian interpolation, so that the update of neighboring Lagrangian points cannot be done independently. This results in a small but non-negligible difference between the Lagrangian point velocity and the fluid velocity on the boundary. Consequently, some streamlines may penetrate in and move out of the solid body, and the boundary layer may separate in advance of the surface.

**Table 1.** Literature review of the key developments in the direct forcing immersed boundary methods.

| IB methods | Literature on IB-NSE | Literature on IB-LBM |
| --- | --- | --- |
| Original | Peskin, 1977[6] | |
| DF | Uhlmann, 2005[7] | Dupuis et al., 2008[14] |
| MDF | Luo et al., 2007[15] | Kang & Hassan, 2011[12] |
| IVC | | Wu & Shu, 2009[16] |
| RKPM | Pinelli, et al., 2010[17] | Li et al., 2016[18] |
| IDF | Kempe & Fröhlich, 2012[19] | Present |
| BTDF | | Present |

In light of this aspect, many improved methods have been developed within the framework of both DF IB-NSE [15, 17, 19] and DF IB-LBM [12, 16, 18], as illustrated in Table 1. The first one is the multi-direct forcing (MDF) technique with multiple forcing iteration steps, as presented by Kempe & Fröhlich [19] and Luo et al. [15] for DF IB-NSE, and Kang & Hassan [12] for DF IB-LBM. In this method, the fluid velocity will gradually approach the particle velocities at the Lagrangian points and the boundary velocity error will be reduced to close to zero, after several or more forcing iteration steps. While the idea is straightforward, the simplicity and accuracy have been improved, and it becomes much more computationally expensive when the system has a large number of particles. The second one is the implicit velocity correction (IVC) method proposed by Wu and Shu [16, 20] in the DF IB-LBM through boundary condition enforcing. In this approach, the boundary force is set as unknown rather than pre-calculated and it is determined after implicitly enforcing the no-slip velocity boundary condition. The third approach is the Reproducing Kernel Particle Method (RKPM) suggested by Favier's group first in the DF IB-NSE [17] and then in the DF IB-LBM [18, 21]. The idea was inspired by Liu et al. [22] by amending the spreading operator with a correct coefficient, which can be considered at each local thickness of the boundary segment where the Lagrangian points are [18], to improve the reciprocity of the interpolation-spreading operations. The last improved method is implicit direct forcing (IDF) presented by Kempe & Fröhlich [19] in the DF IB-NSE. Kempe & Fröhlich [19] proposed this method based on correction of the inequality between the interpolation and the spreading operation in the discrete case, however, they did not implement it for cost reasons. In the above-mentioned latter three methods, the no-slip boundary condition can be satisfied well. However, it relies on resolving a system of linear matrix equations for each particle at each time step, which may become unacceptable for semi-dilute or dense particle-laden flows, especially for 3D and high-Reynolds-number flows. The matrix will become so singular and unsolvable when the neighboring Lagrangian points are too close, when the IB method is coupled with other interface techniques such as the front tracking method[23]. In addition, Park et al. [24] recently proposed a pre-conditioned implicit direct forcing (PIDF) method to take into account the fluid viscosity effect for different Reynolds numbers. However, this makes the DF-IB method become even more complicated and lowers the efficiency for simulation of particle-laden flows, though the matrix can be solved by the conjugate gradient method.

In this work, we first propose an implicit direct forcing (IDF) technique for the IB-LBM coupled method and then transfer it to an explicit direct forcing technique, by enforcing a partition-of-unity condition on the regularized delta function. Based on this approach, we propose a boundary thickening direct forcing (BTDF) method by simplifying the partition-of-unity condition in a subtle way with the minimum error, i.e., uniformly thickening the boundary thickness. Then we suggest optimal values for the boundary thickness for different delta functions according to the numerical accuracy and stability. Furthermore, we will couple various DF-IB methods with LBM for simplicity's sake while not losing the overall generality, for a comparative study of the present BTDF method with the conventional DF [12], MDF [12], IVC [16] and RKPM [18, 21] methods. In addition, the resolution of the Lagrangian points on the particle surfaces is also thoroughly discussed because it has a significant influence on the accuracy, efficiency and stability, which has not been well described in previous papers.

This paper is organized as follows. The derivation of the IDF technique and the proposed BTDF method are presented in Section 2. Theoretical comparisons between the BTDF and other DF methods are given in Section 3. Next, details on the four control parameters in the DF-IB method are presented in Section 4. The simulation results and discussion are presented in Section 5. Finally, the conclusions and a further discussion are given in Section 6.

## 2. Description of the numerical method

### 2.1. Governing equations

For incompressible viscous flows laden with particles, the governing equations for the direct forcing immersed boundary (DF-IB) method, originated by Uhlmann [7], can be expressed as:

$$\nabla \cdot \mathbf{u} = 0 \tag{1}$$

$$\frac{\partial \mathbf{u}}{\partial t} + \mathbf{u} \cdot \nabla \mathbf{u} = -\nabla p + \nu \nabla^2 \mathbf{u} + \mathbf{f} \tag{2}$$

$$\mathbf{f}(\mathbf{x},t) = \int_\Gamma \mathbf{F}(s,t) \delta(\mathbf{x} - \mathbf{X}(s,t)) ds \tag{3}$$

$$\frac{\partial \mathbf{X}(s,t)}{\partial t} = \mathbf{u}(\mathbf{X}(s,t),t) = \int_\Omega \mathbf{u}(\mathbf{x},t) \delta(\mathbf{x} - \mathbf{X}(s,t)) dx \tag{4}$$

where **x**, **u**, $\nu$, $p$ and **f** are the fluid Eulerian coordinates, the vector of fluid velocities, the fluid kinematic viscosity, the fluid pressure normalized with the fluid density and a volume force density term, respectively.

**X** and **F** represent the solid Lagrangian coordinates and the vector of IB force density at the Lagrangian points. Equations (1) and (2) are the mass conservation equation and momentum conservation equation. They are solved in the entire computational domain including the actual exterior domain fluid and the interior domain occupied by the suspended particles. Equations (3) and (4) describe the communications between the fluid phase and the immersed boundary by first interpolating the velocity from the Eulerian grids to Lagrangian points, and then spreading the IB force density calculated at the Lagrangian points back to the Eulerian grids, respectively. The last two equations are only enforced in the local domain $\Omega$ surrounding the immersed boundary $\Gamma$. $\delta(x-X(s,t))$ is the regularized delta function. It is the key to the above-mentioned communication for the DF-IB method.

The classical method applied to advance over time in the incompressible NS equation is the projection method, which is known to cause the Poisson problem, affecting the accuracy of the boundary conditions. In this paper we adopt another fluid solver, the lattice Boltzmann method [12], which directly provides the velocity field without any pressure-correction step while the pressure field can be easily obtained from the density through an ideal state equation. A major problem of the LBM with an external force term in the early stage is the discrete lattice effects, which were eliminated satisfactorily in the Guo et al. [25] model. In this model, the volume force term is introduced with a split forcing LB equation by considering the contribution of the force to both momentum and momentum flux. Consequently, it can correctly recover equations (1) and (2) by a Chapman–Enskog multi-scale expansion with second-order accuracy for either steady flow or unsteady flow, and either uniform external force or non-uniform external force. This was further confirmed by Son et al. [26]. The Guo et al. [25] model has been widely applied in fluid-structure interaction flows.

The split forcing LB equation with single-relaxation time (SRT) from the Guo et al. [25] model can be written as

$$f_\alpha(\mathbf{x}+\mathbf{e}_\alpha \Delta t, t+\Delta t) = f_\alpha(\mathbf{x},t) - \frac{1}{\tau}\left[f_\alpha(\mathbf{x},t) - f_\alpha^{(eq)}(\mathbf{x},t)\right] + F_\alpha(\mathbf{x},t)\Delta t \qquad (5)$$

$$f_\alpha^{(eq)} = \omega_\alpha \rho \left[1 + \frac{3}{c^2}(\mathbf{e}_\alpha \cdot \mathbf{u}) + \frac{9}{2c^4}(\mathbf{e}_\alpha \cdot \mathbf{u})^2 - \frac{3}{2c^2}\mathbf{u}^2\right] \qquad (6)$$

$$F_\alpha(\mathbf{x},t) = \left(1-\frac{1}{2\tau}\right)\omega_\alpha \left[3\frac{\mathbf{e}_\alpha - \mathbf{u}(\mathbf{x},t)}{c^2} + 9\frac{\mathbf{e}_\alpha \cdot \mathbf{u}(\mathbf{x},t)}{c^4}\mathbf{e}_\alpha\right]\cdot \mathbf{f}(\mathbf{x},t) \qquad (7)$$

where $f_\alpha$ and $f_\alpha^{(eq)}$ are the distribution function and the corresponding equilibrium distribution function. $e_\alpha$, and $w_\alpha$ are the discrete velocity in the α-th direction and the corresponding weighting coefficients, respectively. $\tau$ is the dimensionless relaxation time, determined by $\tau=0.5+3\nu/(c^2\Delta t)$. In the D2Q9 model for 2D flows, $w_0=4/9$, $w_{1\sim 4}=1/9$, and $w_{5\sim 8}=1/36$. The discrete velocity vectors $e_\alpha$ are given by

$$[\mathbf{e}_\alpha] = [\mathbf{e}_0, \mathbf{e}_1, \mathbf{e}_2, \mathbf{e}_3, \mathbf{e}_4, \mathbf{e}_5, \mathbf{e}_6, \mathbf{e}_7, \mathbf{e}_8] = c\begin{bmatrix} 0 & 1 & 0 & -1 & 0 & 1 & -1 & -1 & 1 \\ 0 & 0 & 1 & 0 & -1 & 1 & 1 & -1 & -1 \end{bmatrix} \qquad (8)$$

where the lattice speed $c=\Delta x/\Delta t$, and $\Delta x$ and $\Delta t$ are the lattice spacing and time step. The fluid density $\rho$ and velocity $\mathbf{u}^*$ pre-collision can be directly evaluated by taking the zeroth and first moments of

the distribution functions, respectively:

$$\rho = \sum_{\alpha} f_{\alpha} = \sum_{\alpha} f_{\alpha}^{(eq)} \qquad (9)$$

$$\mathbf{u}^{*} = \frac{1}{\rho}\sum_{\alpha} \mathbf{e}_{\alpha} f_{\alpha} = \frac{1}{\rho}\sum_{\alpha} \mathbf{e}_{\alpha} f_{\alpha}^{(eq)} \qquad (10)$$

In addition, the local velocity post-collision will be further corrected by the contribution from the force density $\mathbf{f}(\mathbf{x},t)$. Consider the discrete lattice effect on the forcing term in the lattice Boltzmann method, where the physical fluid velocity during a time step with second order accuracy is the average of the pre- and post-collision velocities [25]:

$$\mathbf{u} = \frac{\mathbf{u}^{pre-collision} + \mathbf{u}^{post-collision}}{2} = \mathbf{u}^{*} + \frac{\Delta t}{2\rho}\mathbf{f}(\mathbf{x},t) \qquad (11)$$

*2.2. Implicit direct forcing IB-LBM*

The IB-LBM method is usually implemented in two main steps [14]:

*Interpolation*: Interpolate the fluid velocity information from the Eulerian grids to the Lagrangian points and then calculate the IB-related force density at the Lagrangian points according to the no-slip boundary condition.

*Spreading*: Spread the obtained force density from the Lagrangian points to the neighboring Eulerian grids and accomplish the solution of the fluid equations at the Eulerian grids with the IB-related force density.

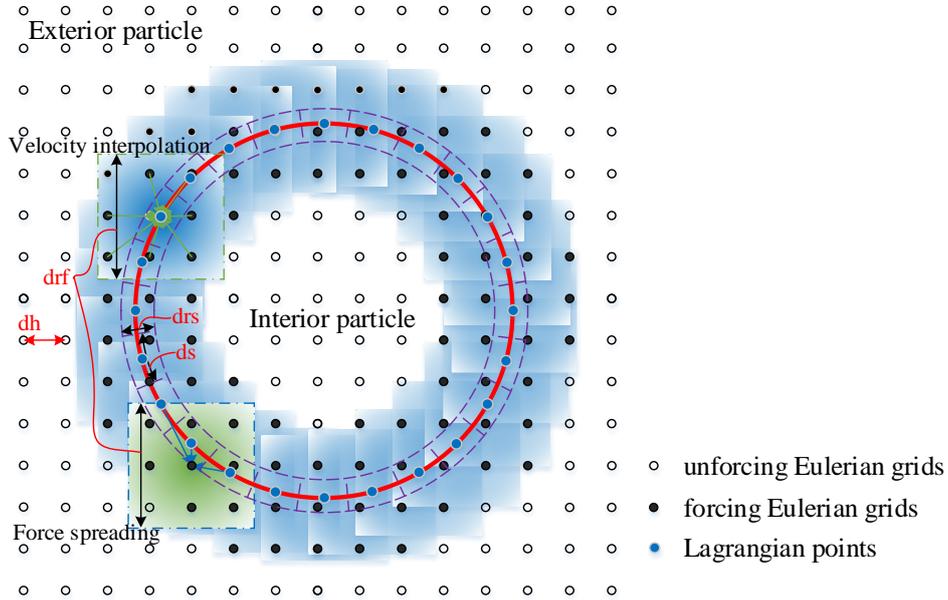

**Fig. 1**. Illustration of the diffusion immersed boundary around the interface of a solid particle. The red heavy circle line represents the physical boundary of a particle. The gradient and transparent square zones, in green or blue, depict the velocity interpolation and forcing spreading range of the regularized delta function. The purple dashed circle band ring and the irregular blue band ring represent the solid forcing shell and the fluid forcing shell, respectively. *dh*-Eulerian grid step (here, *dh*=*D*/9, *D* is the diameter of the circle); *ds*-Lagrangian point step (here, *ds*=*dh*); *drf*-thickness of the fluid forcing shell (here, *drf*=3.0*dh* as the 3-point delta function); *drs*-thickness of the solid forcing shell (here, *drs*=*dh*).

The use of a regularized delta function for the interpolation and spreading manipulation results in an outcome where the sharp interface of a particle is replaced by a thin solid forcing shell overlapping a thin fluid forcing shell, as illustrated in Fig. 1. The solid forcing shell and the fluid forcing shell are according to the ring area encircled by two purple circles and the irregular ring area formed by a set of light blue squares, respectively. On the one hand, the thin solid forcing shell is responsible for the solid forcing and the spreading operation; on the other hand, the thin fluid forcing shell is responsible for the fluid forcing and the interpolation operation. In previous research, the solid forcing shell is always ignored or defaults to its thickness equal to the Eulerian grids step [7]. However, as can be seen in Sections 4 and 5 of this paper, only the right cooperation between these two thin shells can produce a good no-slip and no-penetration boundary condition.

Taking a single particle as an example, there are $N_E$ number of local Eulerian grids surrounding the particle boundary with $N_L$ number of discrete Lagrangian points. The velocity interpolation and force density spreading are expressed by Equations (12) and (13), respectively.

$$D_I \mathbf{u} = \mathbf{U}_b \tag{12}$$

$$\mathbf{f} = D_E \mathbf{F}_b \tag{13}$$

where $\mathbf{U}_b$ and $\mathbf{F}_b$ are boundary velocity and force density, respectively. $D_I$ is the interpolation operator matrix with a dimension of $N_L \times N_E$, and $D_E$ is the spreading operator matrix with a dimension of $N_E \times N_L$.

The matrix elements of $D_I$ and $D_E$ are constructed by

$$D_{I,ij} = \frac{1}{dh^2} \delta_{ij}\left(\frac{x_i - X_j}{dh}\right) \delta_{ij}\left(\frac{y_i - Y_j}{dh}\right) \cdot dh^2 \tag{14}$$

$$D_{E,ji} = \frac{1}{dh^2} \delta_{ij}\left(\frac{x_i - X_j}{dh}\right) \delta_{ij}\left(\frac{y_i - Y_j}{dh}\right) \cdot (drs \cdot ds) \tag{15}$$

where $i$ varies from 1 to $N_L$, and $j$ varies from 1 to $N_E$, respectively. $\delta_{ij}( \cdot )$ is the regularized delta function and will be described in Section 4.3.

Substituting equation (13) into equation (11) and then into equation (12), we can get

$$D_I \mathbf{u}^* + \frac{\Delta t}{2\rho} D_I D_E F_b = \mathbf{U}_b \tag{16}$$

Obviously, the IB force $\mathbf{F}_b$ can be obtained as the following matrix form

$$\mathbf{F}_b = (D_I D_E)^{-1} \frac{2\rho}{\Delta t} (\mathbf{U}_b - D_I \mathbf{u}^*) \tag{17}$$

similar to the expression proposed by Su et al. [27] and Park et al. [24] in IB-NSE, and equation (17) is the implicit direct forcing (IDF) technique for the IB force density solution in IB-LBM. Because $D_I D_E$ is a diagonally dominant and symmetric matrix, the $(D_I D_E)^{-1}$ can be obtained by the conjugate gradient (CG) method [24] and the iterative method [17, 27]. However, it is not appropriate for a complex fluid laden with a large number of particles, especially 3D systems. The $D_I D_E$ is scaled with $O(10^2 \times 10^2)$ and $O(10^3 \times 10^3)$ for 2D particles and 3D particles, respectively. It should be solved on each particle boundary at each time step.

Large numbers of particles will produce larger numbers of $(D_I D_E)^{-1}$ solutions, which finally lead to overwhelming computation. In addition, based on equations (14) and (15), matrix $D_I D_E$ will be a singularity so that it is difficult to solve the inverse matrix $(D_I D_E)^{-1}$ if the Lagrangian points step $ds$ is too small [24, 27].

*2.3. Boundary thickening direct forcing*

On the other hand, the interpolation and spreading operations should satisfy a partition-of-unity condition [17, 28]. That means the IB force density $\mathbf{F}_b$ at the Lagrangian points should be equal to the one obtained by spreading to the Eulerian grids and then interpolation back to the Lagrangian points. The second force interpolation step is

$$\mathbf{F}_b = D_I \mathbf{f} \tag{18}$$

Inserting equation (18) into equation (13) gives

$$\mathbf{F}_b = D_I D_E \mathbf{F}_b \tag{19}$$

As a result, equation (17) can be transferred to an explicit form using equation (19) as:

$$\mathbf{F}_b = \frac{2\rho}{\Delta t}\left(\mathbf{U}_b - D_I \mathbf{u}^*\right) \tag{20}$$

which is the IB force density expression of the original direct forcing method.

The element of $D_I D_E$ is given in detail as

$$\left(D_I D_E\right)_{ij} = \frac{1}{dh^2} \sum_k^{N_E} \delta_{ik}\left(x_k - X_i\right)\delta_{ik}\left(y_k - Y_i\right) \cdot \delta_{kj}\left(x_k - X_j\right)\delta_{kj}\left(y_k - Y_j\right) \cdot drs \cdot ds \tag{21}$$

In almost all the previous research, by adopting $drs=dh$, a matrix $\mathbf{A}$ with a $N_L \times N_L$ dimension is defined as

$$A_{ij} = \left(D_I D_{EI}\right)_{ij} = \frac{1}{dh^2} \sum_k^{N_E} \delta_{ik}\left(x_k - X_i\right)\delta_{ik}\left(y_k - Y_i\right) \cdot \delta_{kj}\left(x_k - X_j\right)\delta_{kj}\left(y_k - Y_j\right) \cdot dh \cdot ds \tag{22}$$

which is the default form of the interpolation and spreading operation used [7, 12].

Inspired by Pinelli et al. [17], the hypothesis is that the boundary thickness varies at different Lagrangian points. The boundary thickness $drs$ with a uniform value is replaced by $\mathbf{drs}=\mathrm{diag}(drs_1, drs_2, \ldots, drs_{N_L})$. Equation (19) can be rearranged as

$$\mathbf{F}_b = \frac{1}{dh} \mathbf{A} \cdot \mathbf{drs} \cdot \mathbf{F}_b \tag{23}$$

By requiring that the boundary thickness $\mathbf{drs}$ be independent of the IB force distribution, the array $\mathbf{drs}$ can be found by solving the linear system below, as indicated by Pinelli et al. [17].

$$\frac{1}{dh}\mathbf{A}\vec{drs}=\vec{1} \tag{24}$$

and

$$\vec{drs}=\mathbf{A}^{-1}\cdot\vec{1}\cdot dh \tag{25}$$

where the $\vec{drs} = (drs_1, drs_2, \ldots, drs_{N_L})^\mathrm{T}$ is a column matrix of the boundary thickness and $1=(1,1,\ldots,1)^\mathrm{T}$.

For simplicity, we propose that the *drs* should be a uniform value considering the physical mean and the uniform distribution of Eulerian grids and Lagrangian points. It can be solved by the average value of the vector $\vec{drs}$ to obtain the minimum mean error as

$$drs = \frac{1}{N_L} \sum_{k=1}^{N_L} drs_k \tag{26}$$

As can be seen in equations (26), (25) and (22), *drs* is decided by the regularized delta function $\delta_{ij}(\cdot)$ and the Lagrangian points step *ds*, which will be depicted in Section 4.4.

Based on the present method, the hydrodynamic force acted on the particles can be simply calculated by

$$\mathbf{F}_p = \sum_{i=1}^{N_L} \mathbf{F}_{bi} \cdot (ds \cdot drs) = \sum_{j=1}^{N_E} \mathbf{f}_j \cdot dh^2 \tag{27}$$

*2.4. Computational procedure*

In the present BTDF method, an appropriate value for the boundary thickness *drs* should be set according to the regularized delta function $\delta_{ij}(\cdot)$ adopted in the beginning of the calculation. In addition, the BTDF method can be coupled with the MDF method with *NF* iterations, though it is unnecessary as depicted in Section 5. In general, the iteration number can be set as *NF*=1.

The overall procedure for the present BTDF method based the IB-LBM method is shown in detail as outlined in Algorithm 1. It can be seen that the calculation algorithm is similar to that of the DF method [12] if *NF*=1, except for *drs*≠*dh*. That means the computational efficiency and memory cost of the present BTDF method is nearly equivalent to the DF method.

**Algorithm 1**: Calculation procedure for the present BTDF method based on IB-LBM.

1. Calculate the fluid velocity $\mathbf{u}^{*,0}$ without the IB forcing in equation (10);
2. Initialize the boundary force density field $\mathbf{F}_b^0$=0, fluid force density $\mathbf{f}^0$=0 and set *k*=0;
3. **while** $k \leq NF$ **do**

    (a) Solve equation (20) for the IB force density $\Delta \mathbf{F}_b$ increment;

    (b) Spread the IB force density $\Delta \mathbf{F}_b$ to the surrounding fluid in equations (13) and (15) for $\Delta \mathbf{f}$;

    (c) Update the boundary force density using $\mathbf{F}_b^k = \mathbf{F}_b^{k-1} + \Delta \mathbf{F}_b$;

    (d) Update the intermediate velocity ($\mathbf{u}^{*,k}$) using $\Delta \mathbf{f}$ with equation (11);

    (e) Update the fluid force density using $\mathbf{f}^k = \mathbf{f}^{k-1} + \Delta \mathbf{f}$;

    (f) k=k+1;

    **end**

4. Calculate the hydrodynamic force acted on the particles in equation (17);
5. Calculate the equilibrium distribution function in equation (6) using the updated fluid velocity;
6. LB collision is by combining the solving of equations (7) and (5);
7. LB streaming

## 3. Comparison with other direct forcing techniques

*3.1. Direct forcing*

In the direct forcing (DF) technique, the IB force density $\mathbf{F}_b$ is explicitly defined as equation (20). The force density is directly calculated from the difference between the intermediate velocity and the desired velocity at the IB boundary. However, the spreading operation is manipulated by a default hypothesis where the boundary thickness is equal to the Eulerian spacing step, as $drs = dh$. The conventional spreading process is

$$D_{EI,ji} = \frac{1}{dh^2} \delta_{ij}\left(\frac{x_i - X_j}{dh}\right) \delta_{ij}\left(\frac{y_i - Y_j}{dh}\right) \cdot (dh \cdot ds) \tag{28}$$

Due to its simplicity, the DF technique was used in the original DF IB-NSE method [7] and the original DF IB-LBM method [12]. It has been widely used in simulations of particle-laden flows. However, it is not appropriate for providing accurate non-slip boundary conditions at the IB boundary because the partition-of-unity condition of equation (19) cannot be satisfied.

*3.2. Multi-direct forcing*

Kang and Hassan [12] proposed a multi-direct forcing (MDF) technique in IB-LBM that determines $\mathbf{F}_b$ iteratively using the following three steps based on the above DF technique (the boundary thickness also defaults to $drs = dh$):

$$\Delta \mathbf{F}_b^k = 2\rho \frac{\mathbf{U}_b - D_I \mathbf{u}^{*,k}}{\Delta t}, \tag{29}$$

$$\mathbf{u}^{*,k+1} = \mathbf{u}^{*,k} + \frac{\Delta t}{2\rho} D_{EI} \Delta \mathbf{F}_b^k, \tag{30}$$

$$\mathbf{F}_b = \sum_{k=1}^{NF} \Delta \mathbf{F}_b^k, \tag{31}$$

This implies that the MDF technique is approximate to a truncated polynomial expansion or Taylor series expansion of the IDF technique in equation (17). It will be equivalent to the IDF technique when $\mathbf{F}_b$ in the MDF method converges. The MDF method is simpler and does not require constructing the $D_I D_E$ matrix and solving its inverse. The memory cost of MDF is equivalent to the present BTDF method and the DF method. It is suitable for the simulation of particle laden flows with a few particles. The disadvantage is that the simulation is overwhelmed by complex fluids laden with a large number of particles, especially 3D particles. The number of Lagrangian points is commonly $O(10^2)$ and $O(10^3)$ for a single 2D particle and 3D particle, respectively. Furthermore, the number of particles may be $O(10^3) \sim O(10^6)$ in the semi-dilute and dense particle-laden flows. Hence, the Lagrangian points would be $O(10^5) \sim O(10^9)$ in total, and the calculation cost for the particle-fluid interaction (IB process) would be considerably higher than that for the solving of the fluid phase.

*3.3. Implicit velocity correction*

In order to directly enforce the no-slip boundary condition at the particle surface in view of the velocity, Wu and Shu [16] proposed an implicit velocity correction (IVC) method. The IVC method can be depicted by an implicit expression as:

$$\mathbf{A}\Delta \mathbf{u}_b = \mathbf{U}_b - D_I \mathbf{u}^*, \tag{32}$$

where $\Delta \mathbf{u}_b$ is the velocity correction of the fluid at the Lagrangian points and matrix $\mathbf{A}$ is defined in equation (22). It can impose the fluid velocity accurately equaling the particle boundary velocity at the Lagrangian points. Then the local fluid velocity is corrected by the intermediate velocity $\mathbf{u}^*$ plus the spreading of the boundary velocity correction $\Delta \mathbf{u}_b$ as

$$\mathbf{u} = \mathbf{u}^* + D_{EI}\Delta \mathbf{u}_b, \tag{33}$$

and the force density at the Lagrangian points and Eulerian grids can be computed by

$$\mathbf{F}_b = \frac{2\rho}{\Delta t}\Delta \mathbf{u}_b, \tag{34}$$

$$\mathbf{f} = \frac{2\rho}{\Delta t}D_{EI}\Delta \mathbf{u}_b, \tag{35}$$

In this method, the boundary thickness defaults to $drs=dh$ in the spreading operator $D_{EI}$, as shown in equations (33) and (35). Actually, the IVC method is equivalent to the IDF technique presented in Section 2.2, as derived in Appendix A. That means the IVC method should still solve a set of algebraic equations as equation (32) or the inverse of matrix $D_I D_E$. The difference is that the IVC method implicitly solves the velocity correction in advance while the IDF technique implicitly solves the force density correction in advance at the Lagrangian points, while the force density is linearly correlated to the velocity as equation (34). Thus, the IVC method will be representative of the IDF technique for comparison and analysis in the rest of this paper.

*3.4. Reproducing Kernel Particle Method*

Inspired by the ideas developed for meshless methods, particularly from the Reproducing Kernel Particle (RKPM) method [22] and from the partition-of-unity concept [28], Pinelli et al. [17] proposed the RKPM method for DF IB-NSE and later expanded it to DF IB-LBM [18, 21]. In the RKPM, the IB force density at the Lagrangian points is explicitly calculated by equation (20) and a correct parameter $\varepsilon_j$ is introduced into the conventional spreading operator as

$$D_{E2,ji} = \varepsilon_j \frac{1}{dh^2}\delta_{ij}\left(\frac{x_i - X_j}{dh}\right)\delta_{ij}\left(\frac{y_i - Y_j}{dh}\right)\cdot (dh \cdot ds) \tag{36}$$

It can be seen that the RKPM is also another form of the IDF technique, whose derivation is given in Appendix B. The difference is that the IDF technique corrects the boundary force density while the RKPM corrects the spreading operator.

The advantages and disadvantages of each method are summarized in Table 2. The DF method is the least accurate method while it is the simplest and least time consuming. The accuracy of the MDF method is related to the computation cost of IB forcing. Higher accuracy requires considerably more time consumption for IB forcing than that in the DF method. The IVC and RKPM methods are similar to the IDF technique. They are the most accurate methods yet they are the most complicated, requiring the largest amount of memory and consuming the most time because they need extra resources to construct the square matrix $D_I D_E$ by matrix $D_I(N_L \times N_E)$ multiplies matrix $D_E(N_E \times N_L)$ and to solve the inverse square matrix $(D_I D_E)^{-1}$. The value of $N_L \times N_E$ will be $O(10^2 \times 10^3)$ for a 2D particle and $O(10^3 \times 10^4)$ for a 3D particle. The construction and matrix inversion solution process is not viable when there are a large number of particles in the system. The present BTDF method combines the advantages of all the above-mentioned methods. On the one hand, it was developed based on the IDF technique using a simplified and approximate manipulation with a minimum mean error. It can obtain considerable accuracy commensurate with the IDF, IVC and RKPM methods, while being much higher than the DF method. It will be proven in Section 5. On the other hand, it is fully equivalent to the DF method with respect to its computational complexity, memory requirements and time consumption. The multiple calculation iterations in the MDF method and matrix related computation in the IDF, IVC, and RKPM methods are avoided.

**Table 2**

The advantages and disadvantages of each DF method for the immersed boundary method.

| Method | Advantages | Disadvantages |
| --- | --- | --- |
| DF [7, 12] | Explicit, simplest, lowest memory requirement, fastest computation speed | Lowest accuracy |
| MDF [12, 15] | Explicit, somewhat simple, lowest memory requirement | Accuracy related to the number of iterations of IB forcing |
| IVC [16] | Implicit, highest accuracy | Large memory and high time consumption to construct the matrix $D_I D_E$ and solve the inverse matrix $(D_I D_E)^{-1}$, instability for small Lagrangian spacing $ds$ |
| RKPM [17, 18] | Implicit, highest accuracy | Large memory and high time consumption to construct the matrix $D_I D_E$ and solve the inverse matrix $(D_I D_E)^{-1}$, instability for small Lagrangian spacing $ds$ |
| Present BTDF | Explicit, simplest, lowest memory requirement, fastest computation speed, high accuracy | Slight loss in accuracy |

**4. Determination of the immersed boundary parameters**

The diffusion strategy for the DF method renders the interface between the fluid and the particle into a thin and porous shell. On the one hand, the Lagrangian shell should be infinitely thin. A thick shell will increase the hydrodynamic diameter of the particles and thus increase the viscosity drag force and pressure drag force. On the other hand, the shell should completely separate the fictitious fluid of the particle interior from the real fluid of the particle exterior. A porous shell will decrease the pressure drag force and increase the viscosity drag force. This is determined by four main parameters related to fluids and solid particles, as shown in Fig.1, which will be described in detail in the analysis below, respectively.

*4.1. Eulerian grids step dh*

The effect of the Eulerian grids step *dh* is straightforward. The accuracy of fluid solving increases and the thickness of the fluid forcing shell is thinned with the decrease in *dh*. Hence, it can decrease the hydrodynamic diameter and decrease the drag force of the particles. However, it cannot effectively reduce the boundary error, because it can neither correct the IB force solving nor correct the spreading operation.

*4.2. Lagrangian points spacing ds*

In the DF-IB method, particle boundaries are approached by a set of Lagrangian forcing points.

Therefore, the distribution and number of these points directly affect the accuracy of this method. In two dimensions, the distribution and number are always investigated using the ratio of *ds*/*dh*, where *ds* is the Lagrangian points spacing. Too high a ratio in *ds*/*dh* will cause fluid leakage and too low a ratio will increase the computational cost, both of which are unnecessary. Uhlmann [7] suggests that *ds*/*dh* =1.0 by default in his original DF-IB method. This means that the Lagrangian points spacing should be equal to the Eulerian grids step. Su et al. [27] and Kang and Hassan [12] adopted *ds*/*dh* = 1/1.5, while Silva et al. [29] used *ds*/*dh*⩽0.9, which means the Lagrangian points spacing should be less than the Eulerian grids step. In this paper, we indicate that *ds* can be further controlled by another parameter, the ratio of *ds*/*drf*, in which *drf* is the diffuse range of the regularized delta function, which will be discussed in Section 4.3.

Actually, the parameter *ds* means the Lagrangian points are impacted by their adjacent points regardless of the strength and manner of that impact. The distance between two adjacent points is *L*=*D*sin(*ds*/*D*). If particle diameter *D*~∞, it will be *L*=*ds*. There exist three situations actually, as illustrated in Fig. 2. The first situation is *L*⩽0.5*drf*, in which both A and B are within the scope of influence of each other. In addition, the Eulerian nodes between them will be influenced by two or more Lagrangian points. If this condition is achieved, the computational accuracy will not be obviously improved by further decreasing the *ds* value. For example, Su et al. [27] used a two-point delta function with *drf*=2.0*dh* to investigate 2D flows past a circle cylinder. They found that the differences in $C_D$ and *St* resulting from the variations in *ds*/*dh* ratios are within 3% when *ds*/*dh* =0.52~0.94. The second situation is 0.5*drf*<*L*⩽1.0*drf*. In this condition, points A and B will not be directly relevant but indirectly relevant through the Eulerian nodes between them, though the nodes are at most influenced by the two points. This will cause a decrease in accuracy. The last situation is *L*>1.0*drf*. It indicates that some Eulerian nodes between the A and B points will be influenced by neither A nor B. In other words, not forcing the nodes will induce fluid leakage through the boundary. This condition should be avoided. In Section 5, we will use the 3-point delta function to further prove this.

Unlike the explicit DF method, *ds*/*dh* can be more constricted in the IDF, IVC and RKPM methods. Too high a resolution of Lagrangian points will result in a Gibbs-like phenomenon [17]. Su et al. [27] show that no singular matrix will be found if the ratio of *ds*/*dh* is greater than 0.5 for the 2-point delta function.

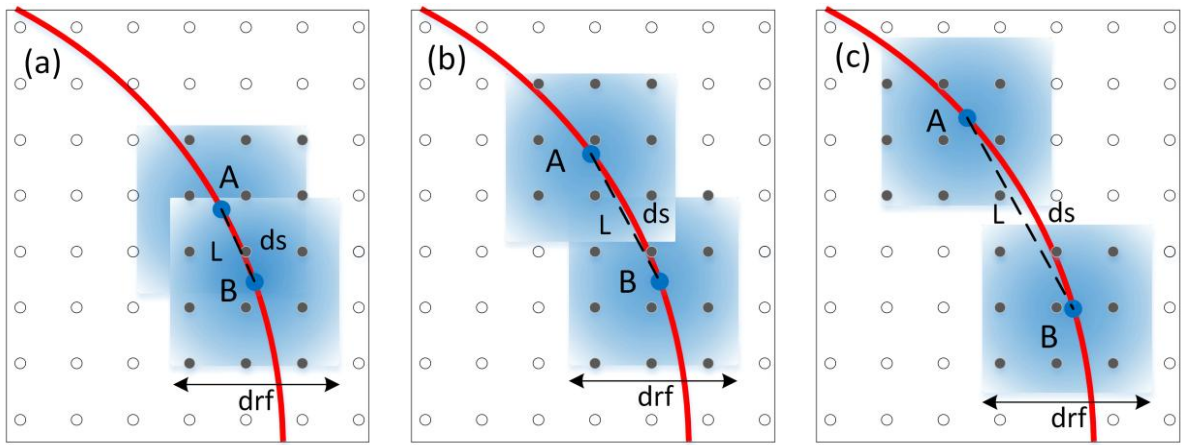

Fig. 2. Illustration of the interaction of adjacent A and B Lagrangian points on a boundary. (a) *L*⩽0.5*drf* ;

(b) $0.5drf \leq L \leq 1.0drf$ ; (c) $L > 1.0drf$ . The thick red lines indicate the particle boundary. The gradient and transparent blue square zones depict the range of action of the regularized delta function.

*4.3. The thickness of the fluid forcing shell drf*

Fluid IB force density acts on the local fluids within the fluid forcing shell and corrects them to satisfy the no-slip condition on the solid–fluid interface. The thickness of the forcing fluid shell depends on the discrete delta function used for velocity interpolation and force density spreading. Three frequently used regularized delta functions will be discussed in this paper. They are the 2-point, 3-point and 4-point delta functions, corresponding to $drf/dh$=2.0, 3.0 and 4.0, respectively.

A 2-point function:

$$\delta(r) = \begin{cases} 0, & |r| > 1, \\ 1-|r|, & |r| \leq 1. \end{cases} \tag{37}$$

A 3-point function:

$$\delta(r) = \begin{cases} 0, & |r| > 1.5 \\ \frac{1}{6}\left(5 - 3|r| - \sqrt{1 - 3(1-|r|)^2}\right), & 0.5 < |r| \leq 1.5 \\ \frac{1}{3}\left(1 + \sqrt{1 - 3|r|^2}\right), & |r| \leq 0.5 \end{cases} \tag{38}$$

A 4-point function:

$$\delta(r) = \begin{cases} 0, & |r| \geq 2, \\ \frac{1}{8}\left(5 - 2|r| - \sqrt{-7 + 12|r| - 4|r|^2}\right), & 1 \leq |r| < 2, \\ \frac{1}{8}\left(3 - 2|r| + \sqrt{1 + 4|r| - 4|r|^2}\right), & 0 \leq |r| < 1 \end{cases} \tag{39}$$

Yang et al. [30] provide a comprehensive test of the various delta functions commonly used in continuous forcing methods. The 3-point function constitutes a good balance between numerical efficiency and smoothing properties. Hence, it will be used in the simulation part of this paper.

*4.4. The thickness of the solid forcing shell drs*

As can be seen in Fig.1, a particle boundary is numerically represented by a ring area shell in two dimensions or a volume shell in three dimensions with a finite thickness *drs*. The IB forcing acts on this shell and drives the particles to move and rotate. Almost all previous studies adopted $drs/dh$ =1.0, as that suggested by Peskin and Uhlmann [7] in their original IBM and original DF-IBM, respectively. In the present BTDF method, we first compute the local thickness of all boundary segments according to each Lagrangian point by combining equations (22) and (25), as shown in Fig. 3. In Fig. 3, the cylinder is centered at an Eulerian node, for example, and the resolution of the fluid and solid is $D/dh$=20 and $ds=dh$ (i.e., the boundary is divided into 124 segments), respectively. It can be seen that the local thickness values decrease from 4- to 3- and 2-point delta functions meaning that the boundary becomes thinner and thinner, i.e., sharper and sharper. In addition, the curve profile of the 2-point delta function is severely oscillating due to the linear interpolation property of this function, while the curve profile of the 3- and 4-point delta functions are more smoothing because these two functions are smoothing. Then these local values are averaged to a uniform thickness value according to the physical meaning with a minimum error by equation (26), as shown in Fig.4.

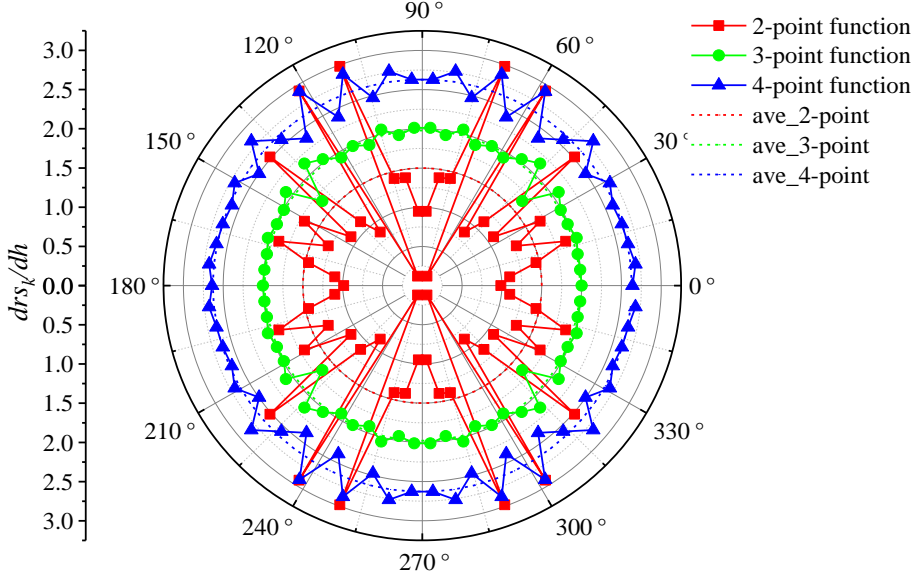

Fig. 3. Local thickness of the boundary segment along the 2D cylinder boundary ($D/dh$=20, $ds/dh$=1.0).

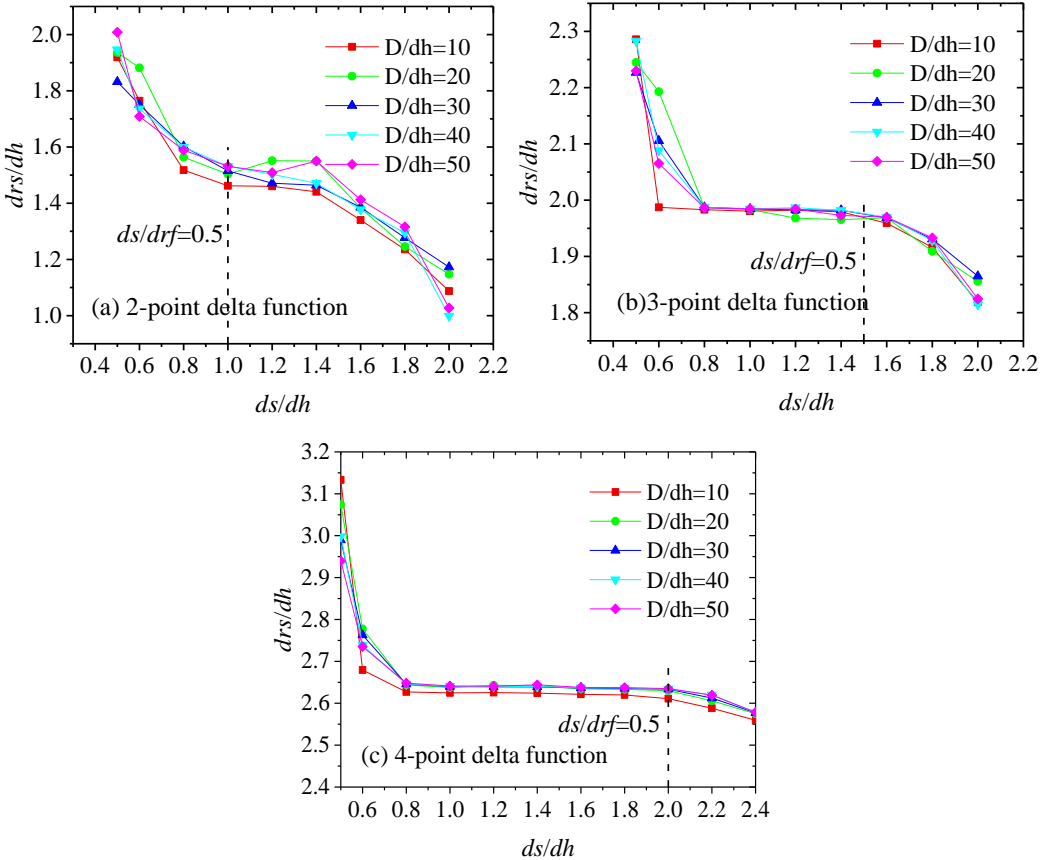

Fig. 4. Average solid forcing shell thickness *drs*: (a) 2-point delta function; (b) 3-point delta function; (c) 4-point delta function.

Figure 4 illustrates the value of *drs/dh* varying with the different delta functions, i.e., *drf/dh*, Eulerian grids solution *D/dh*, and Lagrangian points spacing *ds/dh*. First, the value of *drs/dh* increases with the increase in *drf/dh*. There is not an obvious difference between each *D/dh*. That is to say that the present BTDF method can be applied to any Eulerian grids solution. Second, the value of *drs/dh* decreases with the increase in *ds/dh*. This is because when *ds/dh* increases, the matrix **A** in equations (22) and (24) will converge toward a principal diagonal matrix. That means the determinant of matrix **A** will increase and hence drs/dh decreases. In addition, there are horizontal platforms in Fig.4(b) and Fig.4(c), and the one in the latter is longer than the one in the former. This is because the *drf* action range of the 3-point function and 4-point function is larger than the Eulerian grids step *dh* while *ds* is always scaled with *dh*. That is also why there is no distinct platform in Fig.4(a). Finally, the optimized values of *ds/dh* and *drs/dh* can be pair selected at the platform zone, that is, according to *ds/drf* < 0.5 in Fig. 4. In this paper, we suggest that *drs/dh*=1.4, 1.9 and 2.6 for 2-, 3- and 4-point functions, respectively.

## 5. Simulation results and discussion

To quantify the accuracy and the feasibility of the present BTDF IB method, we simulated several classical benchmark problems to validate the incompressible flow solvers. The 3-point delta function and the corresponding *drs/dh*=1.9 are used in this part.

*5.1. Taylor–Green decaying vortex*

We first adopt the case of a Taylor–Green decaying vortex to evaluate the accuracy of the proposed BTDF method in comparison with the other methods. This test case has been used frequently in previous studies on IB-NSE [7] and IB-LBM [12, 16, 31]. It has the analytical solutions shown below.

$$u_x = -u_0 \cos(\pi x/L)\sin(\pi y/L)e^{-2\pi^2 u_0 t/(\mathrm{Re} L)} \tag{40}$$

$$u_y = u_0 \sin(\pi x/L)\cos(\pi y/L)e^{-2\pi^2 u_0 t/(\mathrm{Re} L)} \tag{41}$$

$$\rho = \rho_0 - \frac{\rho_0 u_0^2}{4 C_s^2}\left[\cos(2\pi x/L)+\sin(2\pi y/L)\right]e^{-4\pi^2 u_0 t/(\mathrm{Re} L)} \tag{42}$$

In this simulation, a stationary circle with diameter *L* is embedded at the center of the fluid computational domain [-*L*, *L*]×[-*L*, *L*], where *L*=1. The exact solutions of equations (40)—(42) at t=0 are imposed as the initial condition. The exact solutions evolving over time are given on the boundary including the outer square boundary, by the non-equilibrium extrapolation method, and the immersed circle boundary, by the present BTDF-IB and other DF-IBs, described in Section 3. The Reynolds number is taken as $Re=u_0 L/v$=10 and the dimensionless relaxation time is set to be $\tau$=0.65, as in [12, 16, 31]. Five different uniform Euler meshes (*N*=10, 20, 40, 80, 160) are used in the simulations. The Eulerian mesh step is $\Delta x=\Delta y=dh=L/N$, the Lagrangian point spacing is *ds=dh*, and the time step is *dt=dh*. The computations are all up to the dimensionless time $T=u_0 t/L$ =1.0. The overall error and boundary error of velocities are evaluated by using the $L_2$-norm error defined as:

$$L_2 - error = \sqrt{\sum_n |u^c - u^a|^2 \Big/ \sum_n |u^a|^2} \qquad (43)$$

where $u^c$ and $u^a$ mean the computational and analytical values, respectively. $n$ is the total number of Eulerian grids in the whole domain for overall error and the number of Lagrangian points on the boundary for boundary error, respectively.

Figure 5 plots the overall error and boundary error versus the Eulerian grids step in the log scale. As can be seen, the error for all these methods is considerable in this case. Furthermore, the convergence order of accuracy is almost second as the slopes of the lines are all about 2.0. This is because the analytical solution is enforced on the circle boundary. Actually, as pointed out by Howell and Bell [32], Roma et al. [33] and Breugem [34], the second-order accuracy in space is due to the velocity field being smooth in this case.

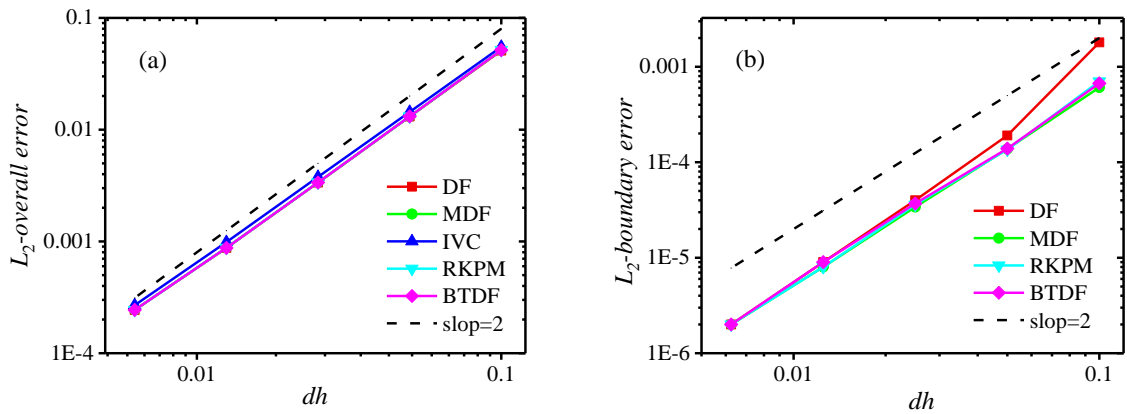

**Fig.5.** Grid convergence study for a Taylor–Green decaying vortex. (a) $L_2$ norms for the overall velocity error and (b) $L_2$ norms for the boundary velocity error.

Figure 6 shows the $x$-velocity magnitude and $x$-velocity gradient plots at $T=1.0$ using the present BTDF method. The velocity values and their gradients at the horizontal middle line and vertical middle line of the computational zone are drawn in Fig. 7. Velocity $u_x$ at the horizontal middle line and $u_y$ at the vertical middle line are not drawn because they equal zero according to equations (40) and (41). It can be clearly seen that both the velocity and velocity gradient are smooth either in the horizontal direction or in the vertical direction.

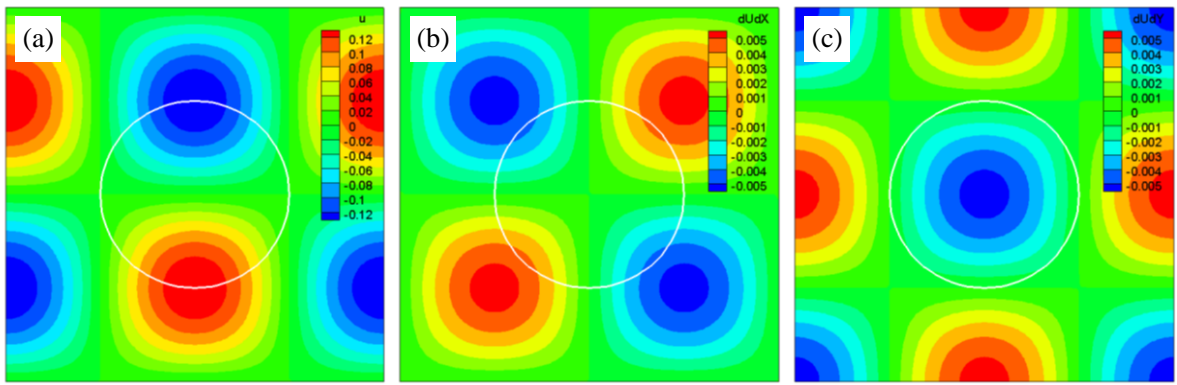

**Fig.6.** Velocity and velocity gradient contour of the Taylor–Green vortex at $T=1.0$ with $L=D=80dh$. (a) $u_x$, (b) $\partial u_x/\partial x$ and (c) $\partial u_x/\partial y$.

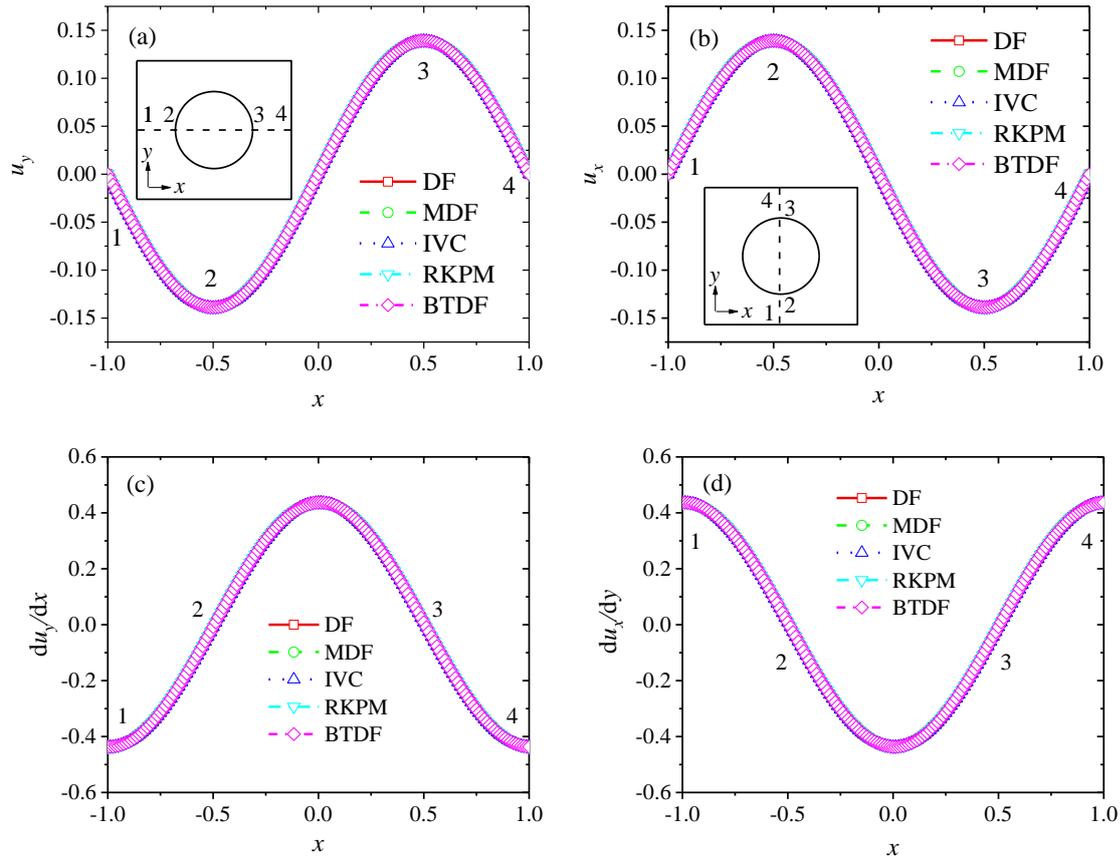

**Fig.7.** Velocity magnitude and gradient at the middle line of the computational zone for a Taylor–Green vortex at $T$=1.0. (a) $u_y$ at the horizontal middle line of $y$=0, (b) $u_x$ at the vertical middle line of $x$=0, (c) $du_y/dx$ at the horizontal middle line of $y$=0 and (d) $du_x/dy$ at the vertical middle line of $x$=0.

*5.2. Flow around a stationary circular cylinder*

The second case is the two-dimensional flow around a stationary circular cylinder, which has a non-smooth velocity field over the boundary [4]. This problem is one of the representative two-dimensional benchmark problems for testing the accuracy of a numerical method in complex geometries. Thus, it has been studied extensively and many comparable results from experimental and numerical investigations can be found. A different kind of flow pattern can be characterized depending on the Reynolds number ($Re = U_\infty D/v$) based on the freestream velocity $U_\infty$, the diameter of the cylinder $D$ and the fluid kinetic viscosity $v$. In this paper, $Re$ ranges from low to moderate values, according to the creeping flow pattern ($Re<1$), steady vortex separating flow pattern ($Re<46$) and unsteady vortex shedding flow pattern ($Re>46$), respectively.

The size of the computational domain is $40D\times 40D$, and the cylinder is located at ($18D$, $20D$). A Dirichlet boundary condition ($u_x/U_\infty=1$, $u_y=0$) is applied at the left inflow boundary. A Neumann condition ($\partial u_x/\partial y=0$) with $u_y=0$ is applied at the top and bottom far-field boundaries, and a convective boundary condition is imposed at the right outflow boundary.

At first, the Reynolds number is set to Re=40. The computational domain is meshed with $1601\times 1601$ ($D/dh$=40) uniform grid points and calculated by MPI parallel strategy. Figure 8 shows the streamlines and

IB fluid force density at the local area around the boundary when flows reach steady states. The results of the conventional DF method versus the different Lagrangian points spacing and the present BTDF method are presented in Fig. 8(a)~(c) and Fig. 8(d), respectively. The symmetric recirculating vortices behind the cylinder are partially truncated to focus on the local area around the cylinder boundary. It can be clearly seen that the streamlines obtained by the conventional DF method enter into boundary through the leading edge and exist outside the boundary through the trailing edge. The decrease in *ds* cannot improve this phenomenon. However, the streamlines obtained by the present BTDF method do not penetrate the cylinder boundary, as illustrated in Fig. 8(d). The exterior fluid is completely isolated from the interior fluid by the boundary. This interior fictitious fluid flows with a pair of large recirculating eddies and behind with a pair of small recirculating eddies at *R*e=40. This is mainly because the no-slip boundary condition is accurately enforced in the present BTDF method, while it is only approximately satisfied in the conventional DF method.

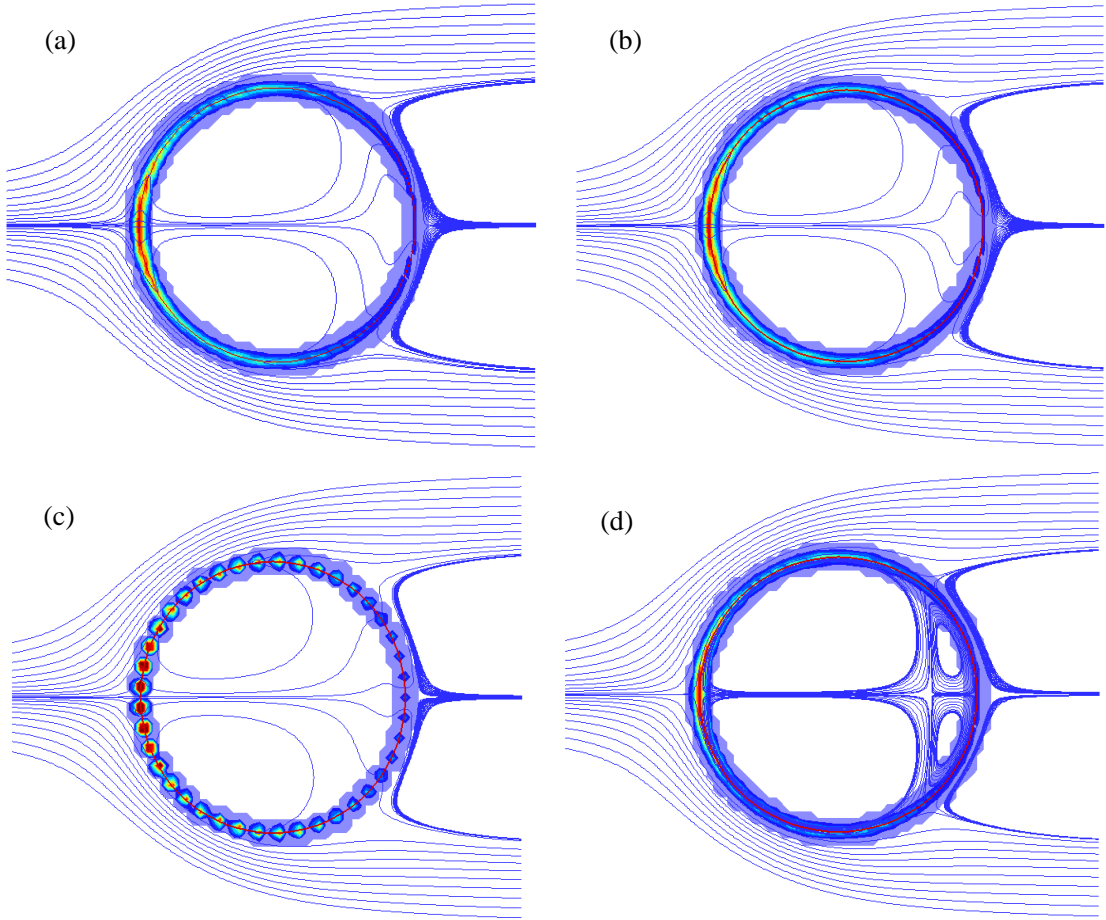

**Fig. 8.** Streamlines (blue line) and IB force density (multicolor contour) on fluid Eulerian grids around the cylinder boundary at Re=40. (a) DF *ds*/*dh*=0.25; (b) DF *ds*/*dh*=1.0; (c) DF *ds*/*dh*=3.0; (d) BTDF *ds*/*dh*=1.0. The thin red circle line represents the real boundary of the cylinder.

The important characteristic parameter associated with this flow is the drag force coefficient defined as

$$C_D = \frac{F_p}{\left(\rho U_\infty^2 D/2\right)}, \tag{44}$$

where $F_p$ is the drag force, which can be calculated in equation (27). $C_D$ is 1.48 and 1.52 in the experimental case and numerical simulation case with the body-fitted grid for $Re$=40 in references [35] and [36], respectively, as shown in Table 3. This means the simulation result is more accurate if it is closer to these two reference values. Since the boundary velocity is zero, the boundary-error to evaluate the no-slip error on the boundary is defined as

$$L_2 - boundary\ error = \sqrt{\frac{1}{n}\sum_n |u_b|^2} \tag{45}$$

Figure 9 presents the time evolutions of the drag coefficient and boundary error versus different values for the $ds/dh$ ratio. The $ds/dh$=1.5 ratio is equivalent to the $ds/drf$=0.5 ratio because the 3-point delta function is used. It proven again that $ds$ cannot obviously improve the boundary error and therefore the drag coefficient, if $ds/dh$≤1.5, as discussed in Section 4. Interestingly, the boundary error decreases when $ds/dh$>1.5, as can be seen on the pink line of $ds/dh$=3.0 in Fig. 9(b). This is because the $ds$ increment has been partially compensated for by the effect of $drs$ on the weight of the spreading operator. However, this improvement is only realized at the Lagrangian points with an isolated local area, as can be shown in Fig. 8(c). The correlation of adjacent points is weaker, and some fluid may leak through the boundary. Hence, it leads to the drag coefficient increase being bad in Fig. 9(a). Fortunately, both the boundary error and drag coefficient are significantly improved by the present BTDF method, as can be seen in the green lines at the bottom of Fig. 9(a) and (b).

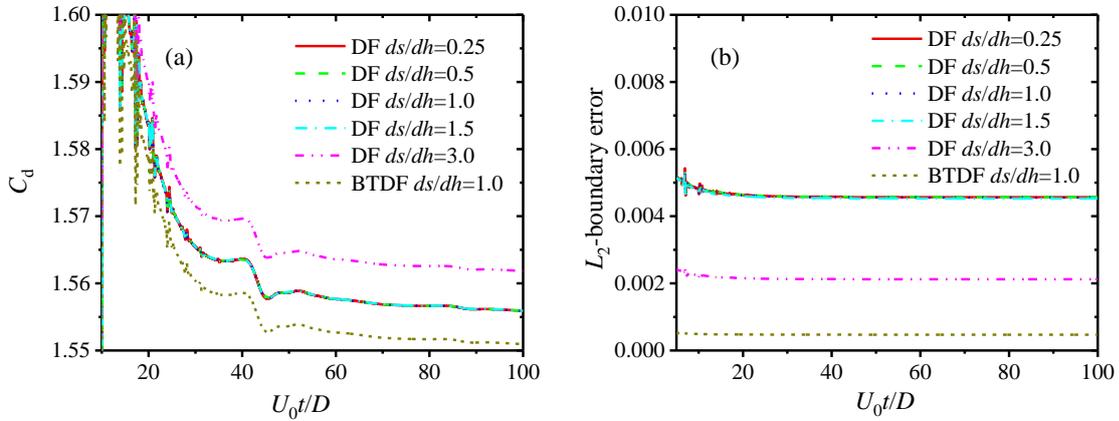

**Fig. 9.** The drag coefficient and $L_2$-error of the boundary velocity affected by the Lagrangian points spacing.

Next, we performed a comparative study of the present BTDF method with the other four methods. Figure 10 shows the normalized normal and tangential velocity on the cylinder boundary. It can be seen that the high non-zero velocities obtained by the DF method have been significantly reduced by the other four methods. The results obtained by the present BTDF method are similar to the MDF, IVC and RKPM methods.

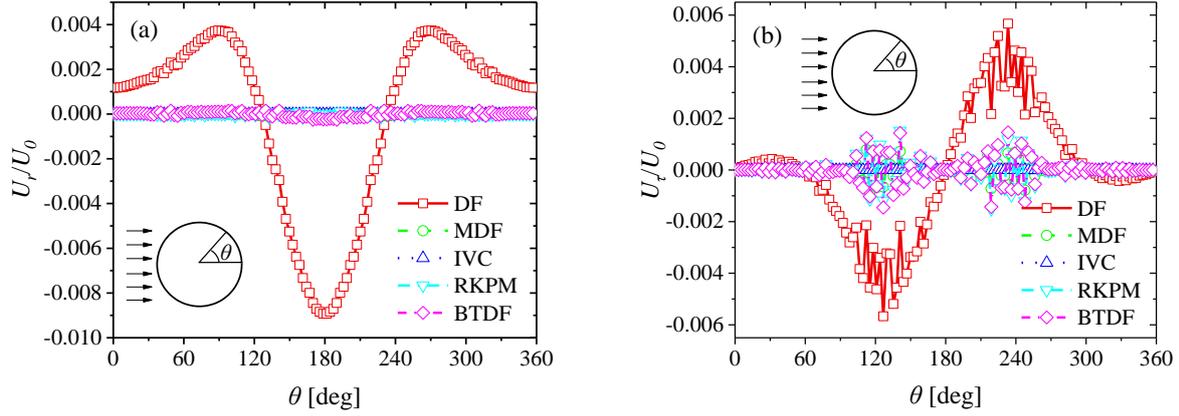

**Fig.10.** Comparison of the normalized velocities on the cylinder surface at Re=40. (a) Normalized radial velocity $U_r$ and (b) Normalized tangential velocity $U\tau$. The number of forcing iterations in the MDF method is 20.

Figure 11(a) depicts the drag coefficient as a function of the grid resolution for various methods. It can be seen that the drag coefficient decreases with the increasing grid resolution. This behavior can be expected for two main reasons. First, the fluid flow field is better resolved with increasing grid resolution. Second, the width of the fluid forcing shell is scaled with the range of the delta function $drf=3.0dh$ and therefore the interface becomes sharper at a higher grid resolution [34]. The drag coefficient $C_D$ obtained by the present BTDF method is similar to the other methods except the DF method and is much lower than the DF method. As can be seen in Table 3, the lower value of $C_D$ is better. Furthermore, the percent error in the drag coefficient $C_D$ as a function of the grid resolution and the various methods is provided in Fig. 11(b). The "exact" reference values of $C_{Dr}$ were obtained from the finest grid resolution $D/dh=160$ in each method. It can be seen that, the slopes of the error lines in the last four methods are about 1.52, which is much higher than the 1.0 obtained with the DF method.

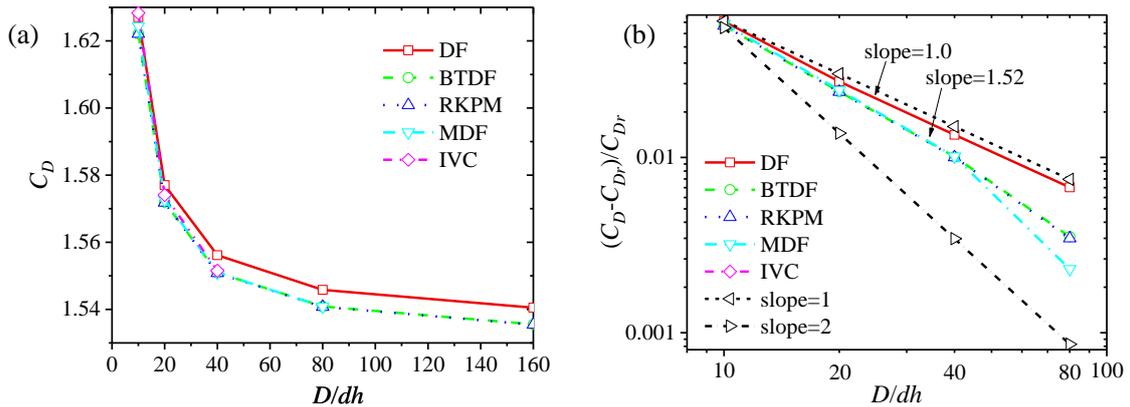

**Fig. 11.** (a) Drag force coefficient $C_D$ as a function of the grid resolution for different direct forcing methods with the Re=40. (b) Percent error in the drag force coefficient $C_D$ as a function of the grid resolution for different direct forcing methods. The error is relative to the value of the drag coefficient $C_{Dr}$ obtained from the finest grid of $D/dh=160$.

To assess the spatial accuracy of the present BTDF method compared with the other methods, the $L_2$ norm error of velocity is redefined as

$$L_2 - over\ error = \sqrt{\frac{1}{n}\sum_n |u - u_{ref}|^2} \tag{46}$$

where $n$ is the number of Eulerian grid points within the 2D×2D control domain surrounding the cylinder, and $u_{ref}$ denotes the "exact" solution obtained in the finest grid ($D/dh=160$).

Figure 12 shows that the velocity convergence rates in the conventional DF method range from 1.17 to 1.36 and from 1.14 to 1.46 for the errors of $u_x$ and $u_y$, respectively. They are improved by the IVC , MDF , RKPM methods and the present BTDF method, as the values range accordingly from 1.18 to 1.56 and from 1.13 to 1.61. They are all smaller than the second-order convergence that is expected for the flow solver without an immersed boundary, which agrees well with the results shown in references [10, 24]. On the other hand, the convergence rates are also smaller than that in the Taylor–Green vortex case. This is because the fluid velocity field is non-smooth near the cylinder boundary, as pointed out by Howell and Bell [32], Roma et al. [33] and Breugem [34].

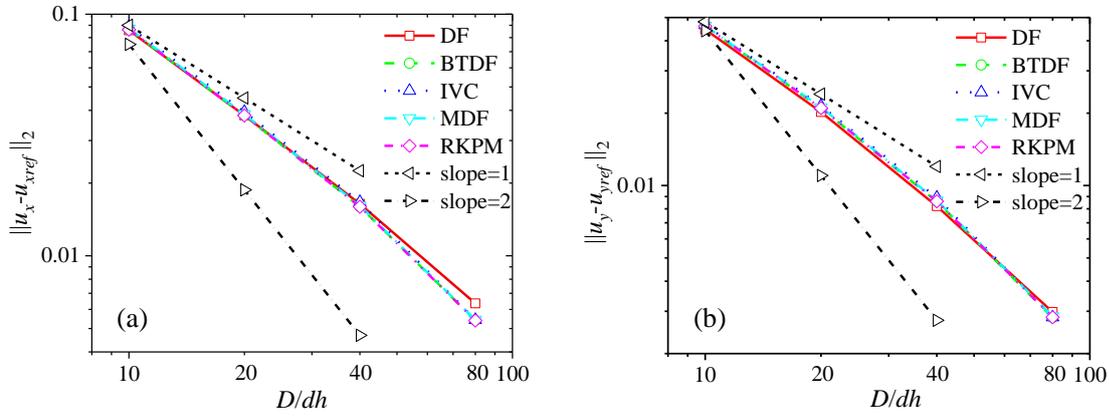

**Fig. 12.** $L_2$ norms for the error in the fluid velocity as a function of the grid resolution for different direct forcing methods. The error is relative to the value of the fluid velocity obtained from the finest grid.

Figure 13 shows the $x$-velocity magnitude and gradient of the $x$- and $y$-direction plots at a stable state using the present BTDF method. It can be seen that there are fluid stagnation zones at the inner and wake length of the cylinder. The velocity gradient is very high at the front edge of the boundary. This will reduce the smoothness of the velocity. The velocity at the horizontal middle line of the cylinder is drawn in Fig. 14(a). In comparison to the conventional DF method, the other three methods with the present BTDF method can effectively reduce the velocity of the front point on the boundary to zero, as can be seen in the small enlarged figure in Fig. 14(a). However, Fig.14(b) shows that there exists a jump in the normal derivative (i.e., gradient in the $x$-direction) at the zone near points A and B over the boundary. It is unlike the smooth velocity gradient of the Taylor–Green vortex case in Fig .7(c) and (d). Furthermore, this is just the origin reducing the velocity convergence rate.

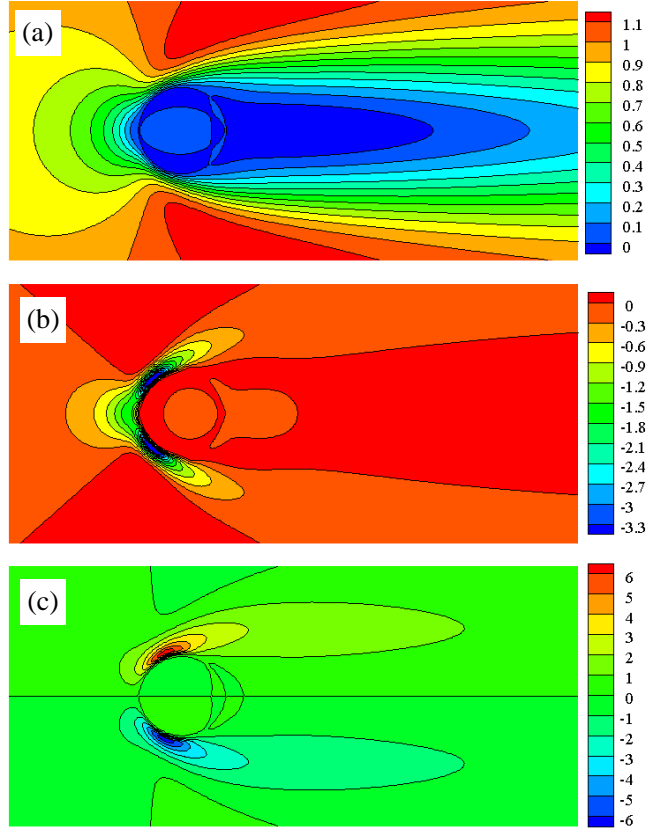

**Fig. 13.** Velocity and velocity gradient contour of the cylinder with $Re$=40 and $D/dh$=40 at a stable state. (a) $u_x$, (b) $\partial u_x/\partial x$ and (c) $\partial u_x/\partial y$.

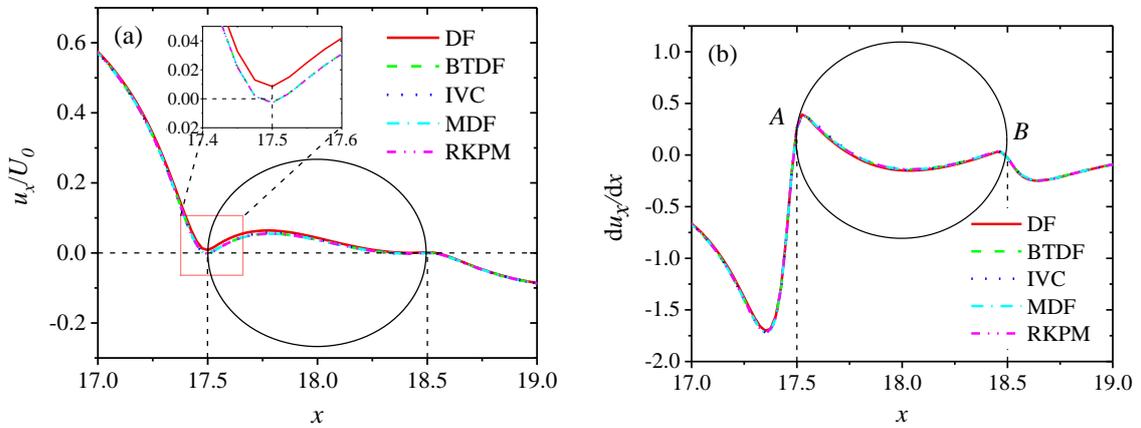

**Fig. 14.** The normalized velocity $u_x$ and its gradient along the horizontal middle line of the cylinder at $Re$=40 and $D/dh$=40. The black circle represents the cylinder boundary.

In order to examine the robustness of the present BTDF method, simulations were performed for different Reynolds numbers. The results were compared with the experimental result [35], numerical result with a body-fitted grid [36] and other numerical results by IB-NSE [24, 37] and IB-LBM [12, 16]. The drag coefficient $C_D$, as well as the wake length $L_w$ for the steady flow and lift coefficient $C_L$ with Strouhal number $St$ for unsteady flow are shown in Table 3. $C_L$ is calculated by the perpendicular component streamwise of $C_D$ in equation (44). Note that $C_D$ and $C_L$ are defined as $a\pm b$ with a mean value of $a$ and a maximum deviation of $b$ at $R$e=200 because the flow is periodically oscillating. The Strouhal number is computed by $St=fD/U_\infty$, where $f$ is the shedding frequency. It can be seen that the present BTDF method accurately predicts not only $C_D$ at a low Reynolds number ($R$e=1 and 40) but also unsteady characteristics such as the shedding frequency and oscillations of $C_D$ and $C_L$ for a moderate Reynolds number ($R$e=200). The calculated values obtained using the present BTDF method agree well with those obtained using a body-fitted grid [36] as well as with other IB methods [12, 16, 24, 37].

**Table 3**

Drag and lift coefficients and Strouhal number for the flow around a stationary circular cylinder ($D/dh$=40).

| References | $R$e=1 | $R$e=40 | | $R$e=200 | | |
|---|---|---|---|---|---|---|
| | $C_D$ | $C_D$ | $Lw$ | $C_D$ | $C_L$ | $St$ |
| Tritton [35] | 11.70 | 1.48 | — | 1.31±0.04 | ±0.65 | 0.20 |
| Choi et al. [36] | — | 1.52 | 2.25 | 1.36±0.048 | ±0.64 | 0.191 |
| Le et al. [37] | — | 1.58 | 2.59 | 1.38±0.040 | ±0.676 | 0.192 |
| Park et al. [24] | 12.00 | 1.54 | — | 1.35±0.04 | ±0.65 | 0.192 |
| Kang & Hassan[12] | — | 1.597 | 2.525 | — | — | — |
| Wu & Shu [16] | — | 1.565 | 2.31 | 1.349 | — | 0.193 |
| DF | 11.266 | 1.556 | 2.42 | 1.39±0.047 | ±0.720 | 0.198 |
| IVC | 11.278 | 1.551 | 2.40 | 1.360±0.044 | ±0.670 | 0.192 |
| RKPM | 11.277 | 1.551 | 2.40 | 1.364±0.042 | ±0.699 | 0.195 |
| MDF (NF=20) | 11.279 | 1.551 | 2.40 | 1.364±0.042 | ±0.699 | 0.193 |
| Present BTDF | 11.277 | 1.551 | 2.40 | 1.364±0.042 | ±0.699 | 0.195 |

**Note:** $C_D$ and $C_L$ for $R$e=200 are defined as $a \pm b$ with a mean value of $a$ and a maximum deviation of $b$.

Figure 15 presents the streamlines around the circular cylinder for low and moderate Reynolds numbers with a grid resolution $D/dh$=40. Visually, the present BTDF method can properly enforce the no-slip boundary condition at the IB surface while preventing the streamlines from penetrating into the IB surface whether the flow is creeping at Re=0.01 or unsteady at Re=200, similar to the MDF method with a NF=20 iterations number. However, the DF method does not accurately predict the velocity fields near the cylinder boundary.

We also performed a comparison study on the effect of the number of forcing iterations $NF$ for predicting the drag coefficients and boundary velocity errors in the BTDF method and the MDF method, as illustrated in Fig. 16(a) and (b). First, there is little effect from $NF$ on the results obtained by the present BTDF method therefore multiple computation iterations may be unnecessary. Nevertheless, there is a larger effect from NF on the decrease in the $C_D$ and boundary error for both MDF and BTDF, especially at the beginning of the five iteration steps. Given a sufficiently large number of forcing iterations, both of the results obtained by these two methods will converge to an identical solution. Second, the results obtained by the present BTDF method without any iteration is better than those obtained by the MDF method with four iteration steps. Hence, the present BTDF method can save on computation time to a much greater degree than the MDF method.

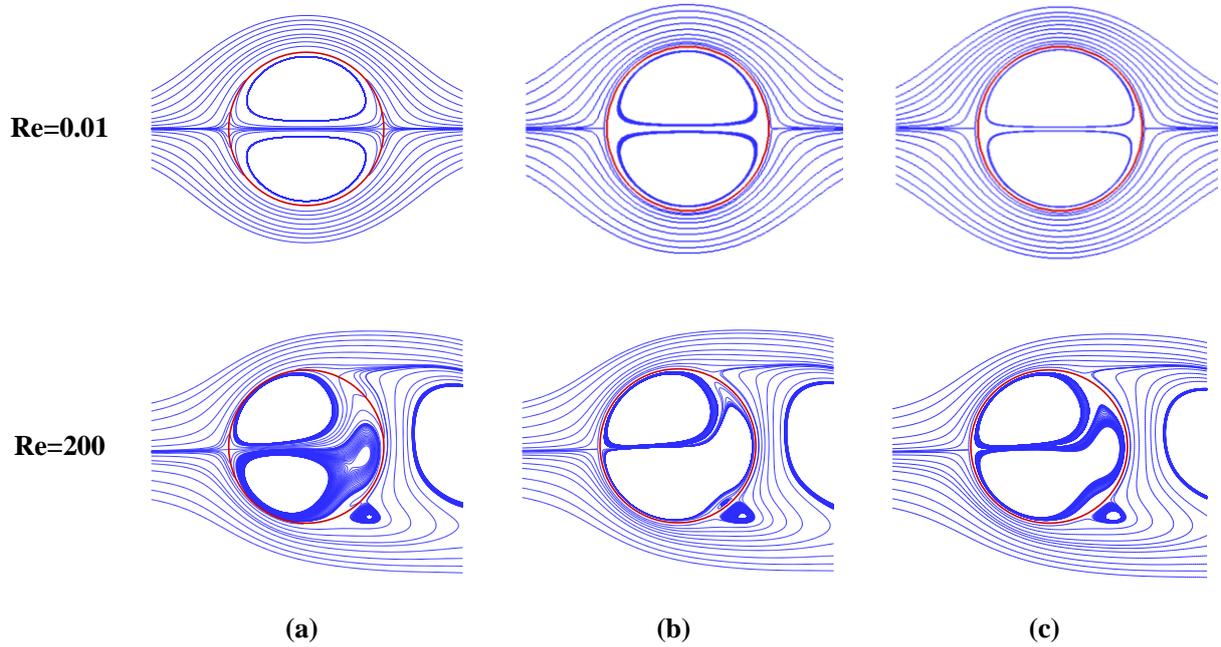

**Fig.15.** Streamlines around a circular cylinder. (a) DF method, (b) MDF method and (c) BTDF method.

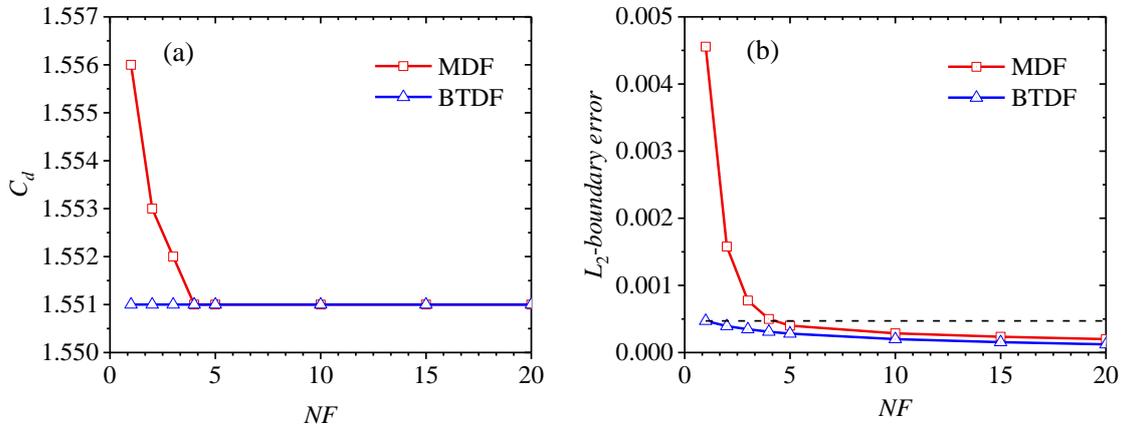

**Fig.16**. BTDF method versus MDF method for flows past a cylinder at $Re$=40 and $D/dh$=40. (a) Drag forcing coefficient $C_D$ as a function of the number of forcing iterations. (b) $L_2$ norm of the boundary velocity error as a function of the number of forcing iterations.

## 6. Conclusion and discussion

We have developed a novel explicit direct forcing immersed boundary method, based on an implicit direct forcing method and a partition-of-unity condition on the regularized delta function for velocity interpolation and force spreading processes. The present BTDF method utilizes an algorithm framework similar to the DF method of Uhlmann [7] used in the IB-NSE method and to that of Kang and Hassan [12] used in the IB-LBM method. The numerical accuracy is increased by a slight thickening of the thickness of the boundary forcing shell, which is usually adopted equal to one Eulerian grid step as employed in previous studies. The no-slip and no-penetration condition on the surface of a particle is better satisfied without any iteration forcing method. In particular, the BTDF method proposed here is comparatively studied with the original DF method [7, 12] and the other three representative improved methods: the MDF [12, 15], IVC [16] and RKPM methods [17, 18]. The main advantages with respect to the other four methods are:

1. With respect to the DF method, the present BTDF method can significantly improve the numerical accuracy to better satisfy the no-slip and no-penetration conditions on the surface of a particle, while the efficiency, memory and computation speed all remain the same.

2. With respect to the MDF method, the present BTDF method can achieve better results than those obtained by the MDF method with more than four iterations. This means that the computation time for the IB forcing in the BTDF method can be four times less than the MDF method. It is very important for the system that the computation load of the IB forcing is considerable or predominately more than that of the fluid flow, such as the semi-dilute and dense particle-laden flows.

3. With respect to the IVC method and RKPM method, the present BTDF method is much simpler, more efficient and stable while the accuracy is similar because the two formal methods are equivalent to the implicit direct forcing method and a complicated coefficient matrix should be solved. However, the present BTDF method is an explicit method simplified from the implicit direct forcing with a minimum error. The present BTDF avoids the large memory usage and inversion solving for the coefficient matrix used in the formal methods at each time step. In addition, the two formal methods are limited to a small Lagrangian points spacing due to the coefficient matrix possibly becoming a singularity to divergence for solving [27], while the BTDF method does not have this limit.

These qualities of the proposed BTDF method have been validated through numerical simulation for 2D creeping, and steady and unsteady flows around a circle cylinder by coupling with the immersed boundary-lattice Boltzmann method. In all cases the numerical results were in excellent agreement with the data in the literature, showing the robustness, accuracy, and efficiency of the proposed methodology.

No tests of the present BTDF method coupled with the IB-NSE method were performed in this study. Nevertheless, the method is inherently general and the only issues that could be faced would be the different fluid solver. A test on 3D particle-laden flows was also not performed. It should be noted that the Lagrangian point will be affected by the adjacent points in eight directions surrounding it in three dimensions, while it is only affected by the front and back in the two directions in 2D. That means the coefficient matrix used in the IVC and RKPM methods will be much larger and denser in 3D and hence more difficult to solve. The BTDF method will see a greater advantage under this condition. The suggested value for the boundary thickness will be provided in research in the near future.

**Acknowledgments**

This work is supported by the National Natural Science Foundation of China (NSFC) (Grant Nos. 51390494, 50936001, 50976042) and the Foundation of State Key Laboratory of Coal Combustion (Grant No. FSKLCCA1802).

**Appendix A: Derivation of the implicit boundary velocity correction (IVC) method [25] from the implicit direct forcing (IDF) technique**

From equation (17), the implicit expression of IB force $\mathbf{F}_b$ is rewritten here

$$\mathbf{F}_b = \left(D_I D_E\right)^{-1} \frac{2\rho}{\Delta t}\left(\mathbf{U}_b - D_I \mathbf{u}^*\right) \tag{A47}$$

First, we confirm that the matrix $D_I D_E$ can be substituted by matrix $\mathbf{A}$ in equation (22) as $drs=dh$ in this method.

$$\mathbf{f} = D_E \left(D_I D_E\right)^{-1} \frac{2\rho}{\Delta t}\left(\mathbf{U}_b - D_I \mathbf{u}^*\right) \tag{A48}$$

It can be seen that, the boundary thickness $drs$ in $D_E$ is offset by the $drs^{-1}$ in $(D_I D_E)^{-1}$. That means the boundary thickness $drs$ can be equal to the Eulerian grids step $dh$ in this method. Therefore, the IB force $\mathbf{F}_b$ can be modified as

$$\mathbf{F}_b = \mathbf{A}^{-1} \frac{2\rho}{\Delta t}\left(\mathbf{U}_b - D_I \mathbf{u}^*\right) \tag{A49}$$

Furthermore, by substituting equation (34) into equation (A49) and multiplying matrix A on both sides, one can find that:

$$\mathbf{A}\Delta\mathbf{u}_b = \mathbf{U}_b - D_I \mathbf{u}^* \tag{A50}$$

Define matrixes $\mathbf{X}=\Delta\mathbf{u}_b$ and $\mathbf{B}= \mathbf{U}_b - D_I\mathbf{u}^*$, the equation (A50) can be rewritten as

$$\mathbf{AX} = \mathbf{B} \tag{A51}$$

This is exactly the expression of the IVC method proposed by Wu and Shu [16]. Thus, it is confirmed that the IVC method is just the velocity version of the IDF technique.

**Appendix B: Derivation of the RKPM from the implicit direct forcing method**

It has been confirmed that the boundary thickness $drs$ can equal the Eulerian grids step $dh$ in Appendix A. Combining equations (13) and (17), we can rewrite the calculation process for the IB force and force spreading as

$$\mathbf{F}_b = \frac{2\rho}{\Delta t}\left(\mathbf{U}_b - D_I \mathbf{u}^*\right) \tag{B52}$$

and

$$\mathbf{f} = D_{EI}\mathbf{A}^{-1}\mathbf{F}_b \tag{B53}$$

Pinelli et al. [17] used a principal characteristic diagonal matrix $\boldsymbol{\varepsilon}$ to substitute $\mathbf{A}^{-1}$. Then, equation (B53) can be modified to the expression of the RKPM as

$$\mathbf{f} = D_{EI}\boldsymbol{\varepsilon}\mathbf{F}_b \tag{B54}$$

where $\boldsymbol{\varepsilon}=\text{diag}(\varepsilon_1, \varepsilon_2, ..., \varepsilon_{NL})$, which is calculated by sum of the corresponding row elements at matrix $\mathbf{A}^{-1}$.